   \documentstyle[aps,prl,twocolumn,epsfig]{revtex}

\begin{document}

   \draft


   \twocolumn[\hsize\textwidth\columnwidth\hsize\csname @twocolumnfalse\endcsname%


   \title{Temperature reconstruction analysis}
   \author{Nicola Scafetta$^{1}$, Tim Imholt$^{2}$, Paolo Grigolini$^{1,3,4}$, and Jim Roberts$^{2}$.}
   \address{$^{1}$Center for Nonlinear Science, University of North Texas,
   P.O. Box 311427, Denton, Texas 76203-1427 }
   \address{$^{2}$Center for Climate Analysis, University of North Texas,
   P.O. Box 311427, Denton, Texas 76203-1427 }
   \address{$^{3}$Dipartimento di Fisica dell'Universit\`a di Pisa and
   INFM, Piazza Torricelli 2, 56127 Pisa, Italy}
   \address{$^{4}$Istituto di Biofisica CNR, Area della Ricerca di Pisa,
   Via Alfieri 1, San Cataldo 56010 Ghezzano-Pisa, Italy}
   \date{\today}
   \maketitle

   \begin{abstract}
 This paper presents a wavelet multiresolution analysis of a time series dataset to study the correlation between the real temperature data and three temperature model reconstructions at different scales. We show that the Mann et.al. model  reconstructs the temperature better at all temporal resolutions. We show and discuss the wavelet multiresolution analysis of the Mann's temperature reconstruction for the period from 1400 to 2000 A.D.E.

   \end{abstract}
   \pacs{} 
   \vspace{0.5cm}
   %
   ] 
   %

\section{Introduction}

The study of past climate trends, or variability is absolutely imperative if any kind of prediction is to be made about the future of our global environment, \cite{jonesagain}.  With the goal of understanding how our climate has changed in the past, several different models have been developed from various Paleoclimatology proxies, \cite{stokstad}.  These proxy records, located all over the globe, are inherent in reconstruction temperature trends from periods before instrumentation records are available.  These longer temperature records are necessary due to the short span of actual temperature records taken via instrumentation.  The longer the time series of temperature data available the better a long term prediction of future temperature behavior can be made.  These proxy data sets include information recovered from the records of tree rings, pollen samples, ice cores, corals, marine sediments as well as others.  The models constructed, in order to be more accurate, take into account many different proxies in determining past climate trends.  This paper makes a comparison of these climate models with instrumentation readings in a new way.  The results show that the model produced by Mann \emph{et al.} \cite{mann} gives a slightly better fit to known temperature values than several of the other paleoclimate models.

\section{Temperature models}

There are three climate models compared in this paper, Fig. 1.  The comparisons are made to known temperature anomaly data taken via instrumentation and reported by Jones \emph{et al.} \cite{temperature}.  This data set is largely considered as the most accurate global temperature data set known.  The models compared are those of Mann \emph{et al}, \cite{mann}, \cite{manndata} Briffa \emph{et al}, \cite{briffa}, \cite{briffadata} and Jones \emph{et al.} \cite{jones}, \cite{jonesdata}  These models all contain at a minimum data from the years 1400 A.D.E. to 1980 A.D.E. (All dates from this point are to be considered A.D.E.)

The Jones \emph{et al.} \cite{jones}, \cite{jonesdata} dataset was generated by averaging 17 temperature reconstructions from both hemispheres.  These sites, once averaged together, form a dataset 1000 years long with a temporal resolution of 1 year.  The proxy data types utilized were tree rings, ice cores, corals and historical documents.  The Mann \emph{et al} \cite{mann}, \cite{manndata} data was a reconstruction from sites globally and extends from the years 1400-1980.  The reconstruction was performed utilizing proxy data from tree rings, ice core, ice melt, long historical records (from Bradley and Jones \cite{historical}), coral, and long instrumentation records.  The Briffa \emph{et al} \cite{briffa} \cite{briffadata} dataset extends from 1400-1994 and was a reconstruction utilizing only tree ring data.

\section{Wavelet multiresolution analysis}

Wavelet analysis  \cite{percival} is a new powerful method to analyze time series. 
Wavelet Transform makes use of scaling functions, the wavelets,     that have
the characteristics of being localized in space and in frequencies. These functions must integrate to zero and the square of them must integrate to unity.
A scaling
coefficient $\tau$ characterizes a wavelet. The length $2\tau$ measures the width of the wavelet. Two typical
wavelets widely used in the continuous wavelet transform are the Haar wavelet 
and the Mexican hat wavelet \cite{percival}.
The Haar wavelet is defined by 
\begin{equation}\label{haarwds}
^{\left(H\right)}\tilde{\psi}_{\tau,t}(u)\equiv \left\{
\begin{array}{ccc}
-1/\sqrt{2\tau},          & ~~t-\tau<u<t    \\
1/\sqrt{2\tau},          & ~~t<u<t+\tau   &    \\
 0,         & ~~otherwise   &.  \\
\end{array}\right.
\end{equation}
The Mexican hat wavelet is the second derivative of a Gaussian.
The length $2\tau$ defines the scale analyzed by
the wavelet.
 Given a signal $\xi(u)$, the Continuous Wavelet Transform  is defined by
\begin{equation}\label{cwtdhjj3}
W(\tau, t)=\int\limits_{-\infty }^{\infty } \tilde{\psi}_{\tau ,t}(u)~\xi(u)~du~.
\end{equation}
The original signal can be recovered from its Continuous Wavelet Transform via
\begin{equation}\label{invcwtxu9}
\xi(u)=\frac{1}{C_{\tilde{\psi}}} \int\limits_{0}^{\infty} \left[ \int\limits_{-\infty }^{\infty } W(\tau,t) ~\tilde{\psi}_{\tau,t}(u)~dt \right] ~\frac{d\tau}{\tau^2}~.
\end{equation}
The double integral of Eq. (\ref{invcwtxu9}) suggests that the original signal may be decomposed in ``continuous details'' that depend on the scale coefficient $\tau$. However, it is not easy to handle  the results of the Continuous Wavelet Transform because of the continuous nature of the decomposition. There exists a discrete version of the wavelet transform, the Maximum Overlap Discrete Wavelet Transform (MODWT), which is the basic tool needed for studying time series of $N$ data via wavelet. In the book of Percival and Walden, ``{\it Wavelet Methods for Time Series Analysis} \cite{percival}, the reader can find all mathematical details.      For the purpose of this paper, it is important to have in mind only one of the important properties of the MODWT, the  Wavelet Multiresolution Analysis (WMA). It is possible to prove that given an integer $j_0$ such that $2^{j_0}<N$, where $N$ is the number of the data, the original time series represented by the vector ${\bf X}$ can be  decomposed in:
\begin{equation}\label{decomw}
{\bf X}=S_{J_0} + \sum _{j=1}^{J_0} D_j~, 
\end{equation}
with 
\begin{equation}\label{decomrel}
S_{j-1}= S_{j} + D_j~.
\end{equation}
The detail $D_j$ represents changes on a scale of $2\tau=2^{j}$, while the smooth $S_{J_0}$ represents averages of a scale of $2\tau= 2^{J_0}$.  To better appreciate the WMA, Fig. 2 shows the WMA of the temperature  in the years 1860-2000. The analysis is done by using the Daubechies {\it least asymmetric}  scaling wavelet filter (LA8) \cite{percival}. LA8 wavelet look similar to the Mexican hat but they are asymmetric.   We plot the WMA for $J_0=4$.  Fig. 2a shows the temperature (dashed line) and the smooth average $S_4$ (solid line). Figs. 2b-e show the details $D_4$,  $D_3$, $D_2$, $D_1$. Detail $D_4$ shows the presence of irregular oscillations with a period of almost 20-22 years. Detail $D_3$ shows the presence of more regular oscillations with a period of almost 10-11 years. Details $D_2$ and $D_1$ show the fluctuations of the temperature at scale $2\tau=2^2=4$ and $2\tau=2^1=2$ years respectively. According to Eqs. (\ref{decomw}) and (\ref{decomrel}), the sum of all four details and the smooth average $S_4$ give the original signal. The smooth average $S_3$ is given by $S_4+D_4$. The smooth averages $S_2$ and  $S_1$ can be obtained by summing first $D_3$ and then $D_2$ to $S_3$  respectively.

\section{Multiresolution correlation analysis}
The use of the Multiresolution Correlation Analysis via wavelet is a simple procedure. We decompose the temperature data and the three temperature reconstructions by  Briffa, Jones and Mann by using 
WMA introduced in the previous section. Then we evaluate the linear correlation coefficient $r$ between the real temperature data and each of the reconstructions for each smooth average and each detail. For pairs of quantities $(x_i; y_i)$; $i= 1, ..., N$, the linear correlation
coefficient $r$ is given by the formula
\begin{equation}\label{corrcoeff}
r=\frac{\sum _{i} \left(x_i-\overline{x} \right)\left(y_i-\overline{y} \right)}
{\sqrt{\sum _{i} \left(x_i-\overline{x} \right)^2}\sqrt{\sum _{i} \left(y_i-\overline{y} \right)^2}}~,
\end{equation}
where, as usual, $\overline{x}$ is the mean of the $x$, $\overline{y}$ is the mean of the $y$.
The value of $r$ lies between  -1 and 1, inclusive. It takes on a value of 1, termed
``completely positive correlation,'' when the data points lie on a perfect straight line
with positive slope, with x and y increasing together. The value 1 holds independent
of the magnitude of the slope. If the data points lie on a perfect straight line with
negative slope, y decreasing as x increases, then r has the value  -1; this is called
``completely negative correlation.'' A value of $r$ near zero indicates that the variables
x and y are uncorrelated. When a correlation is known to be significant, $r$ is one conventional way of summarizing its strength. Because we want to determine which temperature model reproduces better the real temperature at each scale, the best model is the one that gives a linear correlation coefficient $r$ closest to 1. Table 1 shows  the correlation coefficient between the real temperature data and each of the reconstruction for each smooth average and each detail.

 Table 1 shows clearly that Mann's model is the best one in all cases, because its linear correlation coefficient is the closest to 1 for each analysis. Briffa's model is the worst one. However, if we analyze the details, Briffa's model is better than Jones' model. This is information that a simple look at Fig. 1 could not determine.   Moreover, the linear correlation coefficients for the smooth averages are closer to 1 than the correlation coefficients concerning the details. This suggests that the models are made in such a way to reproduce better the smooth averages than the details. Finally, we observe that the worst correlation is for the details $D_{3}$ in all cases. This is very interesting because, as shown in Fig. 2c,  the details $D_{3}$ represents changes on a scale of $2\tau=2^3=8$ years that evidence the 10-11 years temperature periodicity that is connected with the 11 years solar cycle. The low value of $r$ for the details $D_{3}$ suggests that the models give results uncorrelated to real data because of a random shifting.  Fig 3 shows the comparison between the details $D_{4}$, $D_{3}$, $D_{2}$ and $D_{1}$ from Wavelet Multiresolution Analysis of the temperature reconstruction by Mann's model  (solid line) and the real temperature (dashed line) during the year period (1856 - 1980). Figs. 3a, 3c and 3d show that the details $D_{4}$, $D_{2}$ and $D_{1}$ of the temperature reconstruction afforded by Mann's model are satisfactorily correlated to the corresponding properties of the data concerning real temperature.  As for $D_{3}$, illustrated in Fig. 3b, we notice a kind of random shift between reconstruction and real temperature curve.  In some time regions the maxima and minima of the reconstruction curve occur earler, and in other time regions they occur later.  However, the period and teh amplitude of the 10-11 year cycle seem to be the same in both cases.

\section{Mann's  model analysis}

In the previous section we proved that the Mann's  model is the one that produces the most accurate temperature reconstruction of the real temperature data.  The quality of agreement between temperature reconstruction and real temperature becomes unsatisfactory only for the wavelet detail $D_{3}$, a detail of some importance since it refers to the time scale $2\tau=2^3=8$ years, which is very close to the 10-11 year solar cycle.              Figs. 4 show the WMA of the Mann's temperature reconstruction in the period  1400-1980 for $J_0=4$. Fig. 4a  shows the Mann temperature model (dashed line) and the smooth average $S_{4}$ (solid line).The detail $D_{4}$ of Fig. 4b shows that the fastest changes on a scale of $2\tau=2^4=16$ years happened during the periods 1485-1530, 1615-1660, 1685-1710, 1760-1790 and 1810-1840. The detail $D_{1}$ of Fig. 4e shows that the changes on a scale of $2\tau=2^1=2$ year during the period 1700-1980 look stronger than those during the period 1400-1700. Further research is needed to determine whether the effect observed in the detail $D_{1}$ is natural or is due to a lack of precision of the data used to reconstruct the temperature. Finally, Fig. 5 shows the smooth average $S_{3}$ (solid line) for Mann's temperature reconstruction in the period  1400-1980. The dashed line is the real temperature during  the years 1856-2000. The smooth average $S_3$ is the best smooth reconstruction of the temperature that can be done. In fact in Section  4, we proved that the correlation between the real temperature and the temperature reconstruction fails for  details $D_{3}$. Therefore,  it is not realistic to include those details, as well as the details $D_{2}$ and $D_{1}$, in a plausible smooth reconstruction of the temperature.    However, in accordance again with the results of Section 4, the details $D_{2}$ and $D_{1}$ may be considered good enough for studying the changes on the scales of $2\tau=2^2=4$ and $2\tau=2^1=2$ years, respectively. Fig. 5 shows that the coldest  period occurred during the years 1455-1475. From 1500 to 1920 the temperature had little fluctuation, only slightly more than 0.1 degree. There was a warm period from  1765-1778. Since 1920 the temperature had the highest growth in the last six centuries, almost 0.7 degree. The temperature decreased slightly during the period  1945-1975 and then it started to increase again during the last 25 years.

\section{Conclusion}

The conclusions clearly show that, at a minimum, wavelet analysis is a useful technique for comparison of these datasets.  The superior dataset appears to be the Mann \emph{et al.} \cite{mann}, \cite{manndata}.  Future work will include utilizing this wavelet analysis technique in attempting to improve the existing models or in constructing a new paleoclimate model so that a better understanding of current and future climate behavior may be obtained.


   \onecolumn

\newpage

\begin{figure}

\epsfig{file=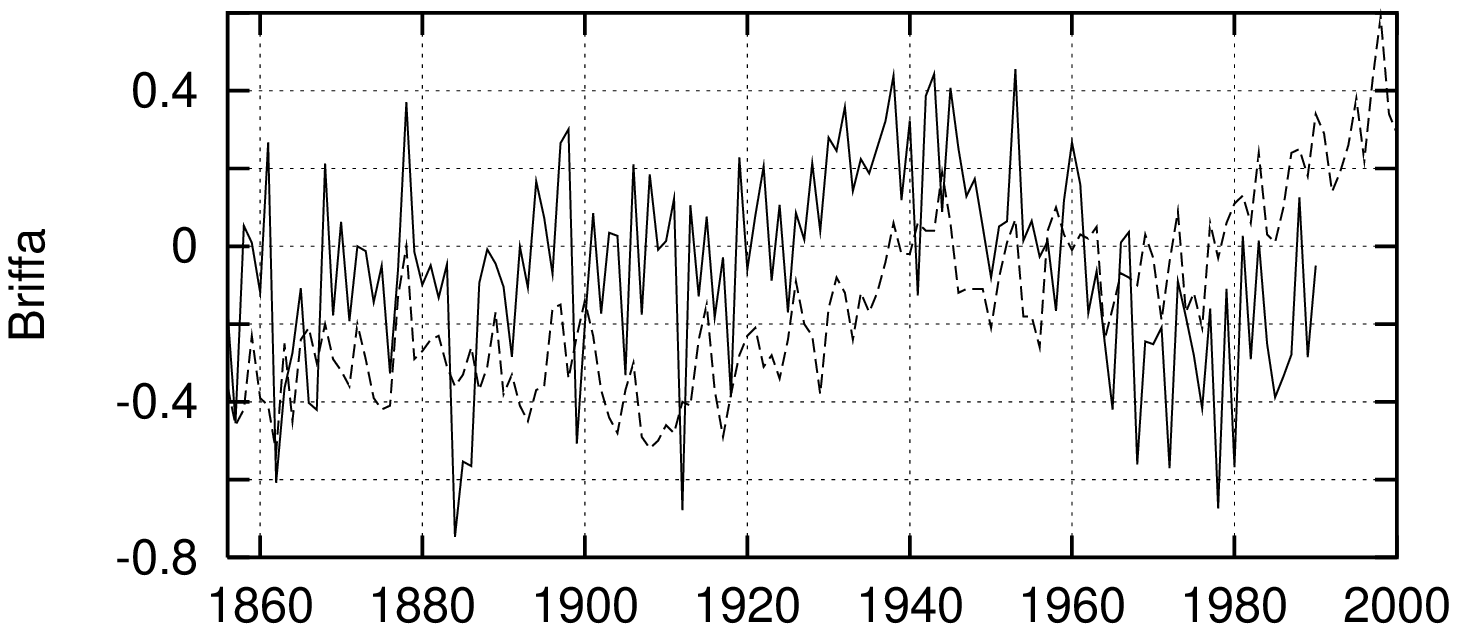,height=18cm,width=6cm,angle=-90}

\epsfig{file=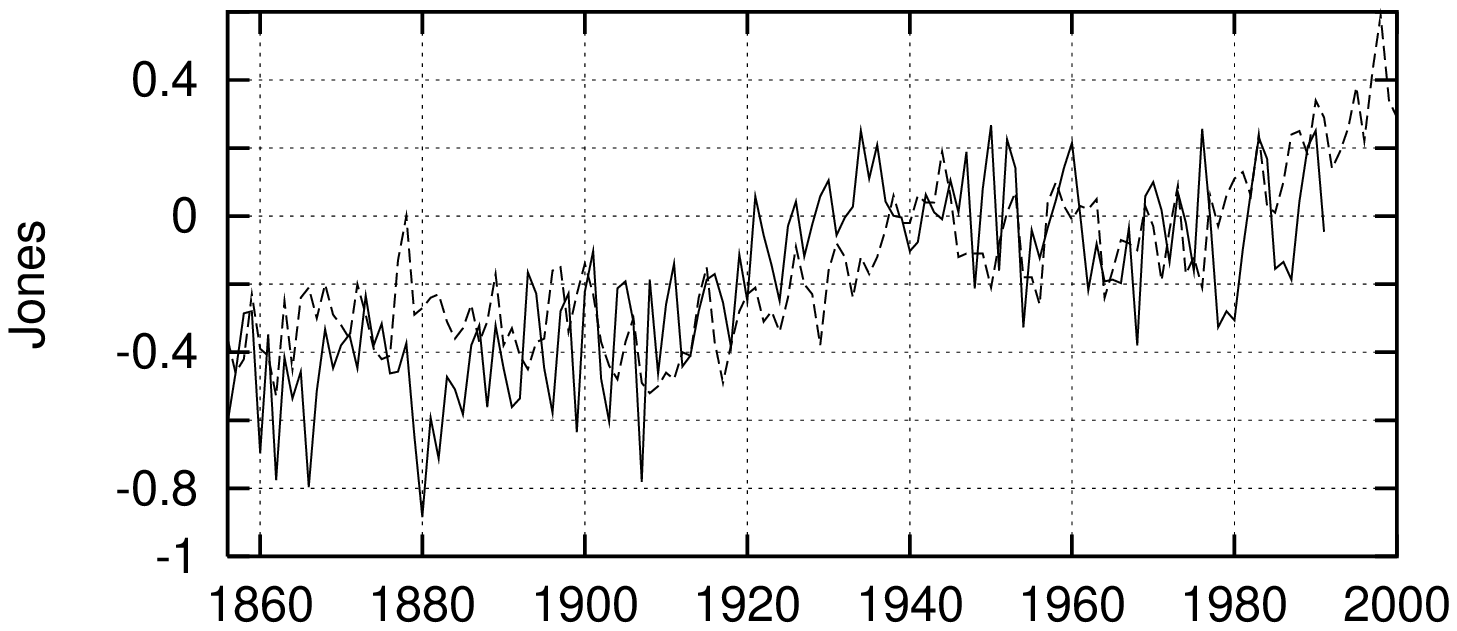,height=18cm,width=6cm,angle=-90}

\epsfig{file=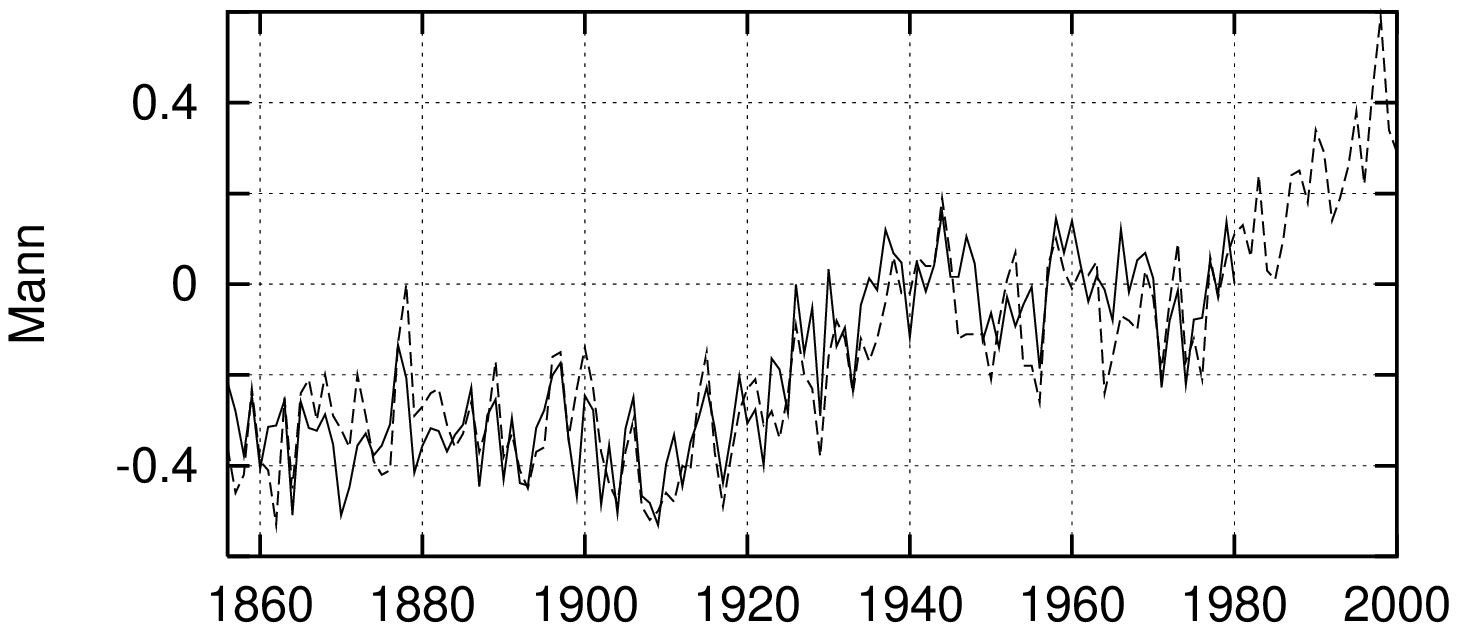,height=18cm,width=6cm,angle=-90}

\caption{Temperature reconstruction (1860:2000): Briffa, Jones and Mann models. The dashed line is the real temperature.}

\end{figure}

\newpage

\begin{figure}

\epsfig{file=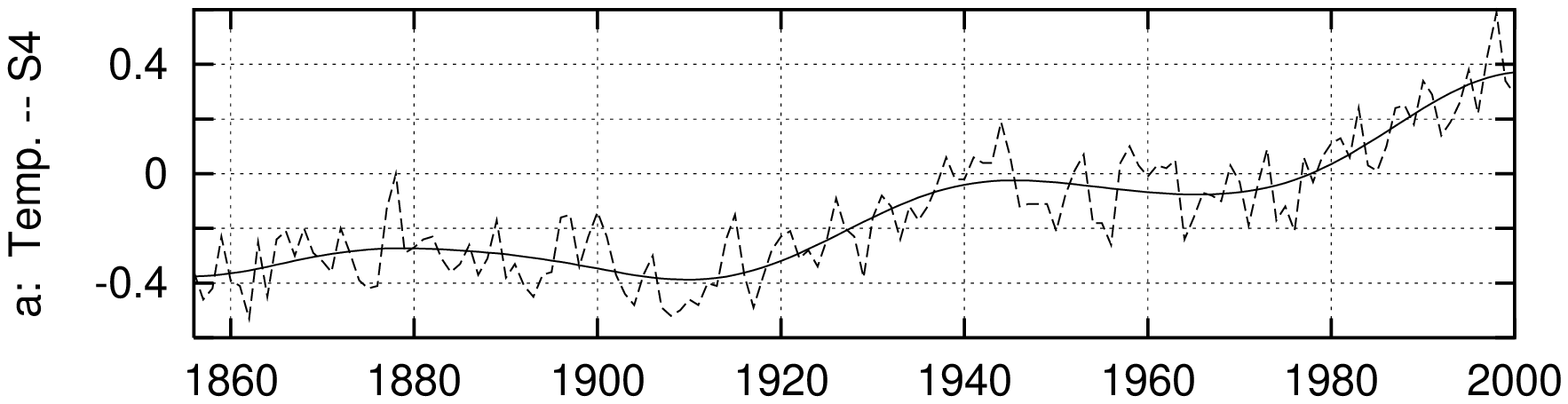,height=18cm,width=4cm,angle=-90}

\epsfig{file=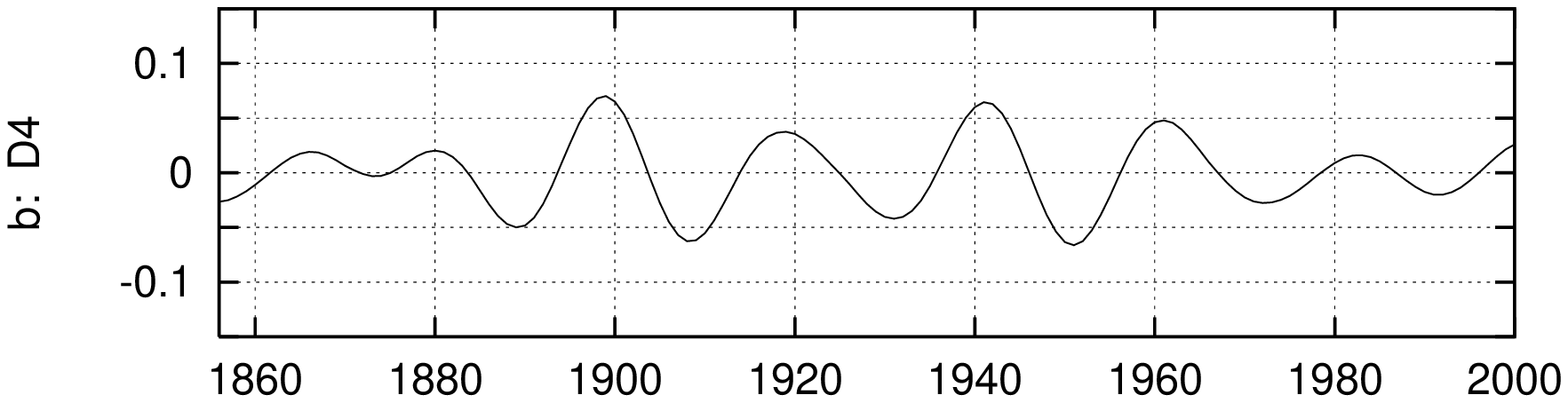,height=18cm,width=4cm,angle=-90}

\epsfig{file=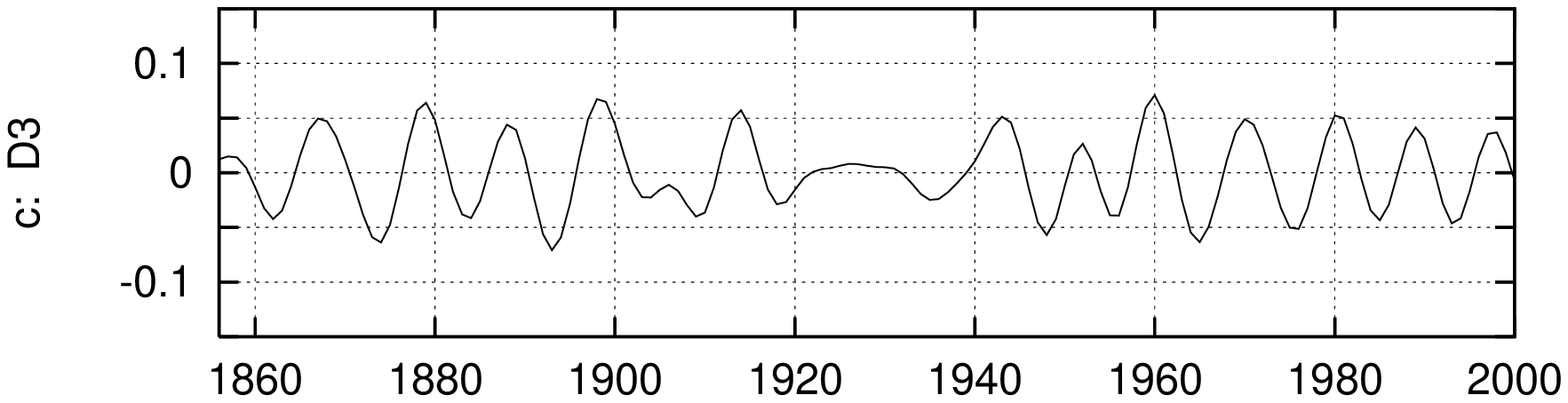,height=18cm,width=4cm,angle=-90}

\epsfig{file=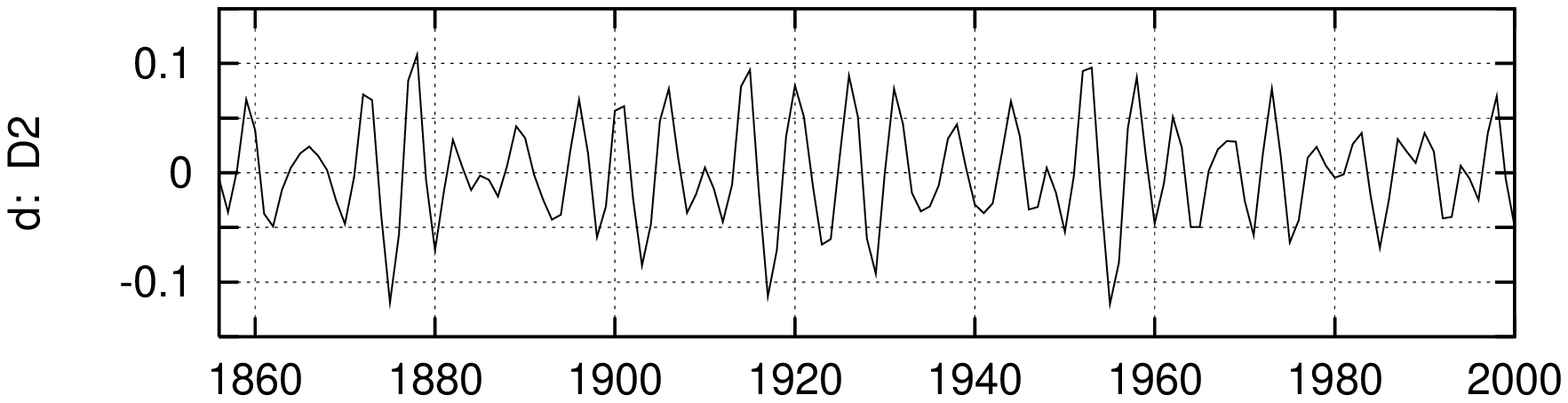,height=18cm,width=4cm,angle=-90}

\epsfig{file=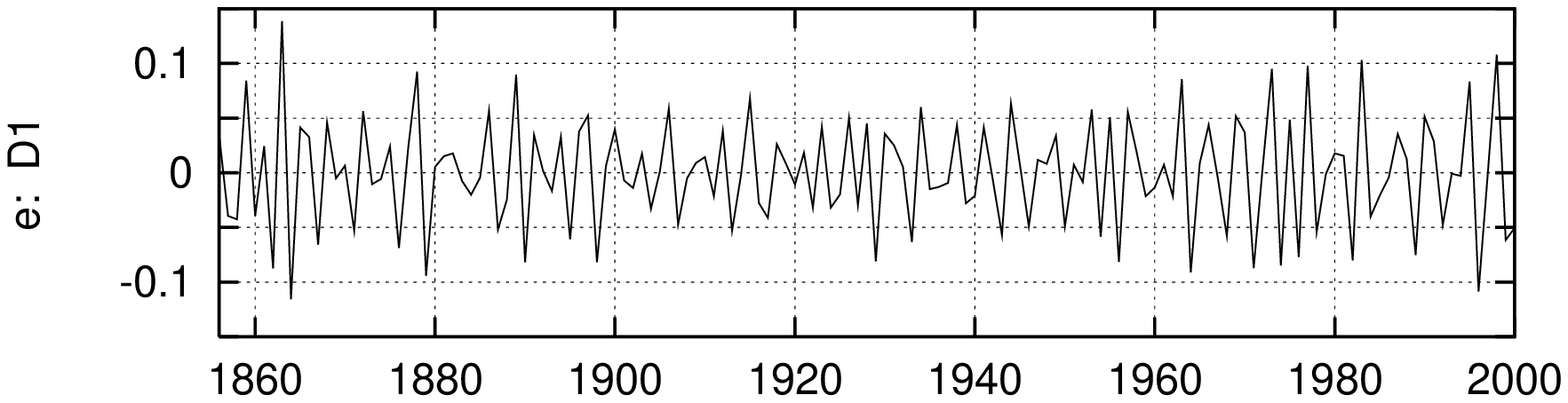,height=18cm,width=4cm,angle=-90}

\caption{Wavelet Multiresolution Analysis $J_0=4$ of the temperature. Fig. 2a shows the temperature (dashed line) and the smooth average $S_4$ (solid line). Figs. 2b-e show the details $D_4$,  $D_3$, $D_2$, $D_1$.}

\end{figure}

\newpage

\begin{figure}

\epsfig{file=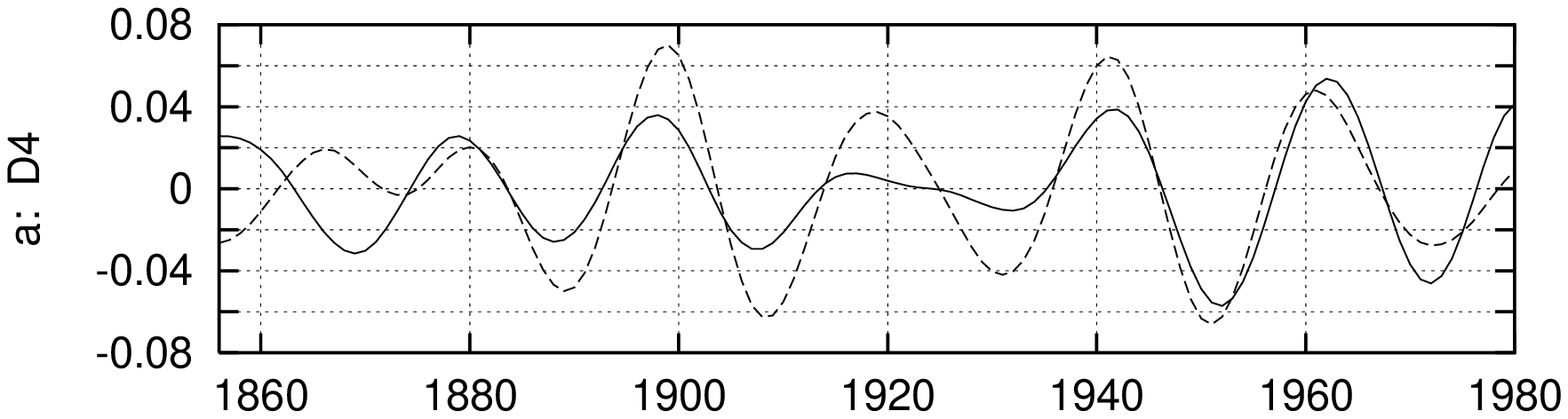,height=18cm,width=5cm,angle=-90}

\epsfig{file=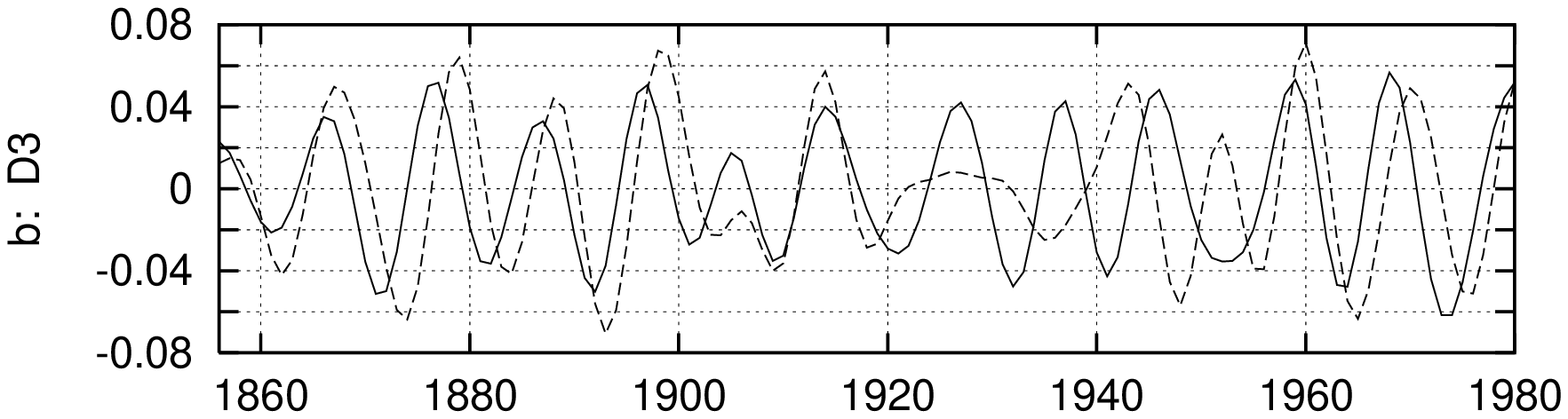,height=18cm,width=5cm,angle=-90}

\epsfig{file=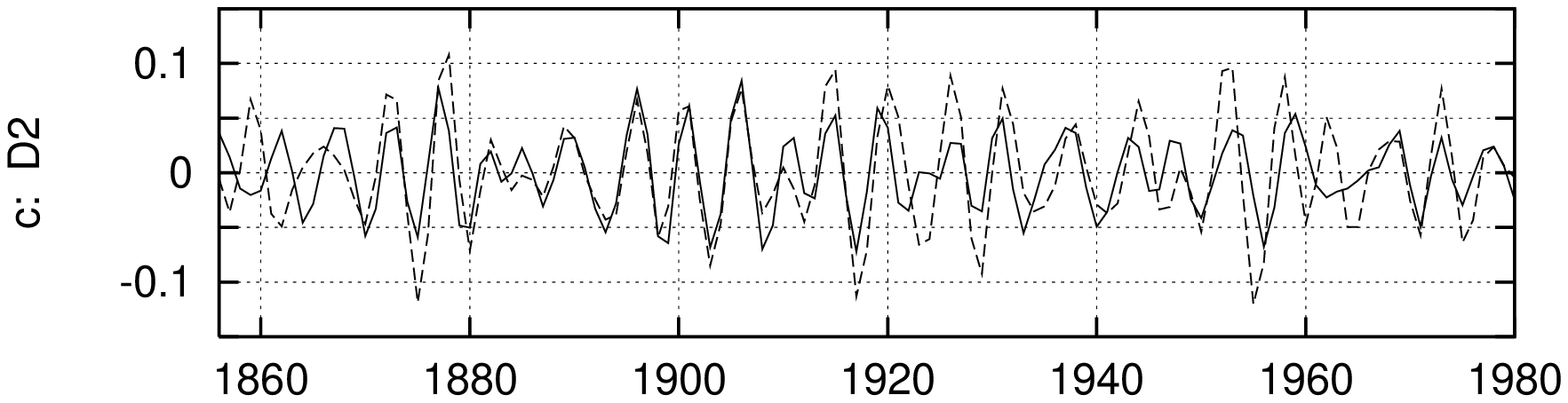,height=18cm,width=5cm,angle=-90}

\epsfig{file=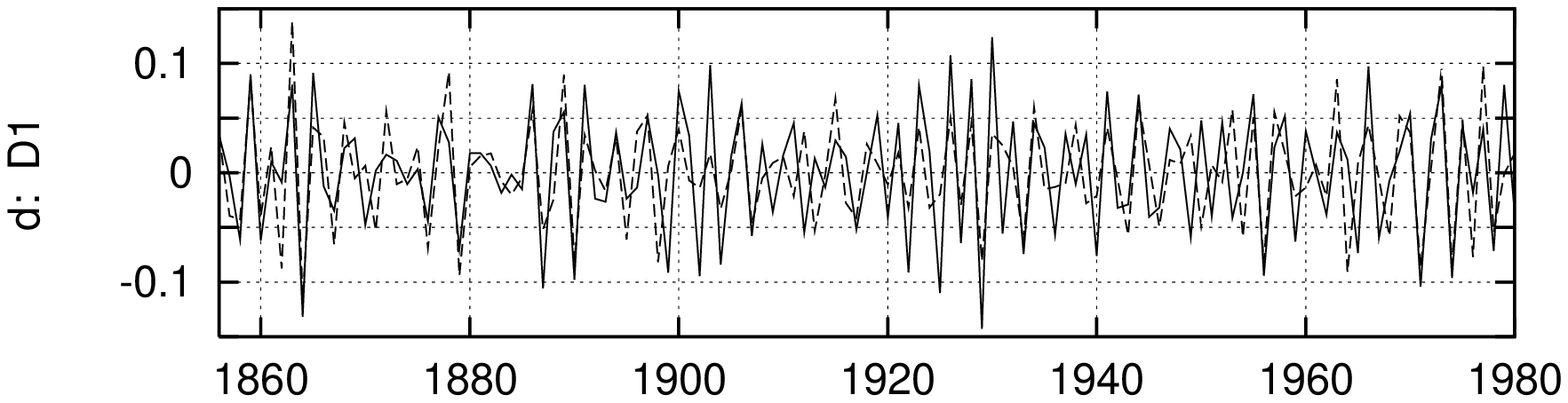,height=18cm,width=5cm,angle=-90}

\caption{Comparison between Temperature reconstruction according to Mann's model (solid line) and real temperature (dashed line).  We show the details $D4$, $D3$, $D2$ and $D1$ obtained in both cases with the method of Wavelet Multiresolution Analysis: year (1856:1980).}

\end{figure}

\newpage

\begin{figure}

\epsfig{file=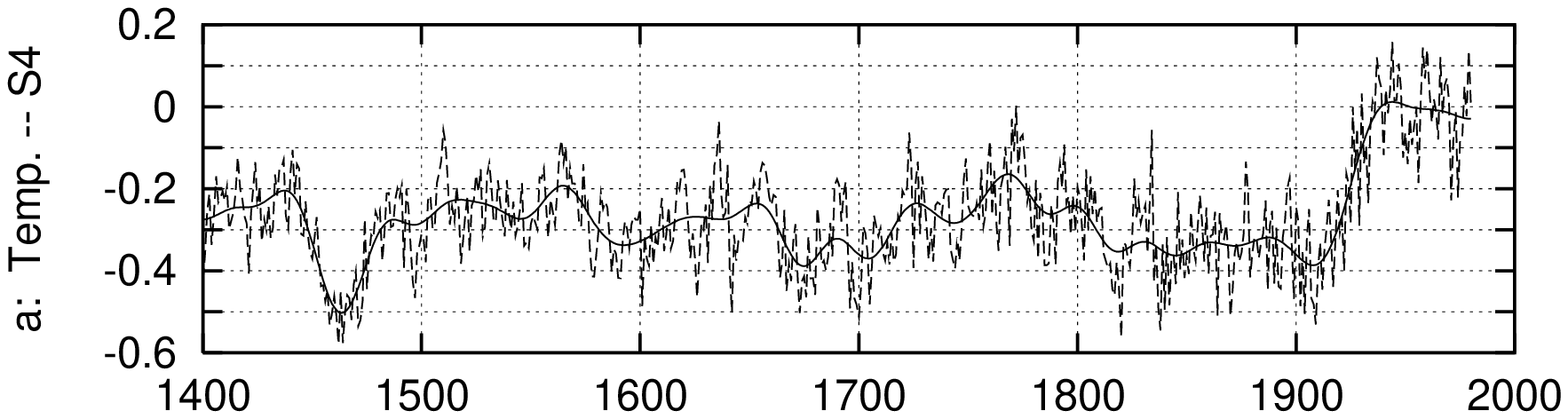,height=18cm,width=4cm,angle=-90}

\epsfig{file=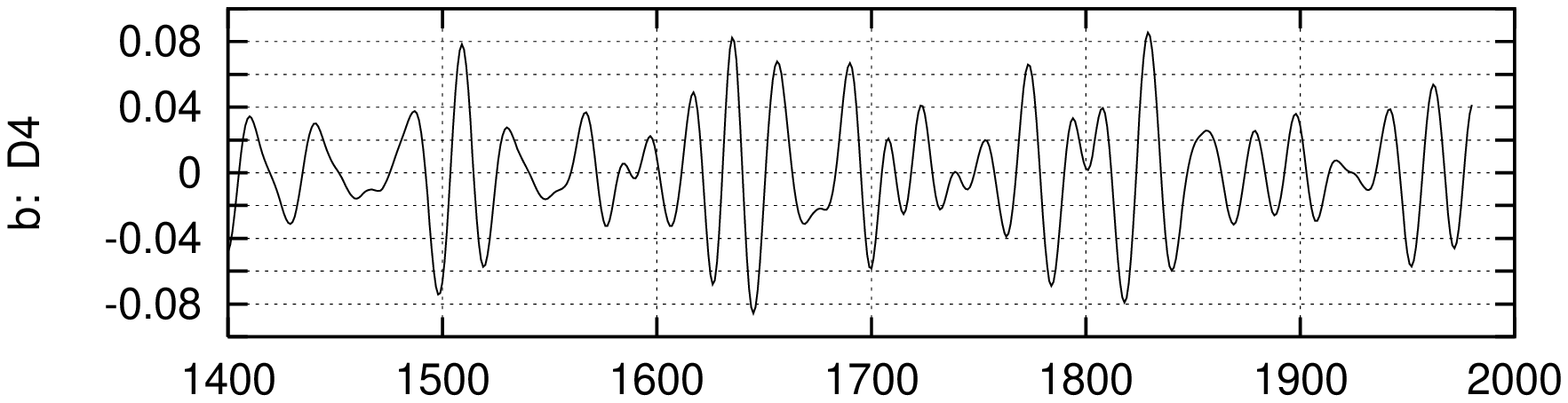,height=18cm,width=4cm,angle=-90}

\epsfig{file=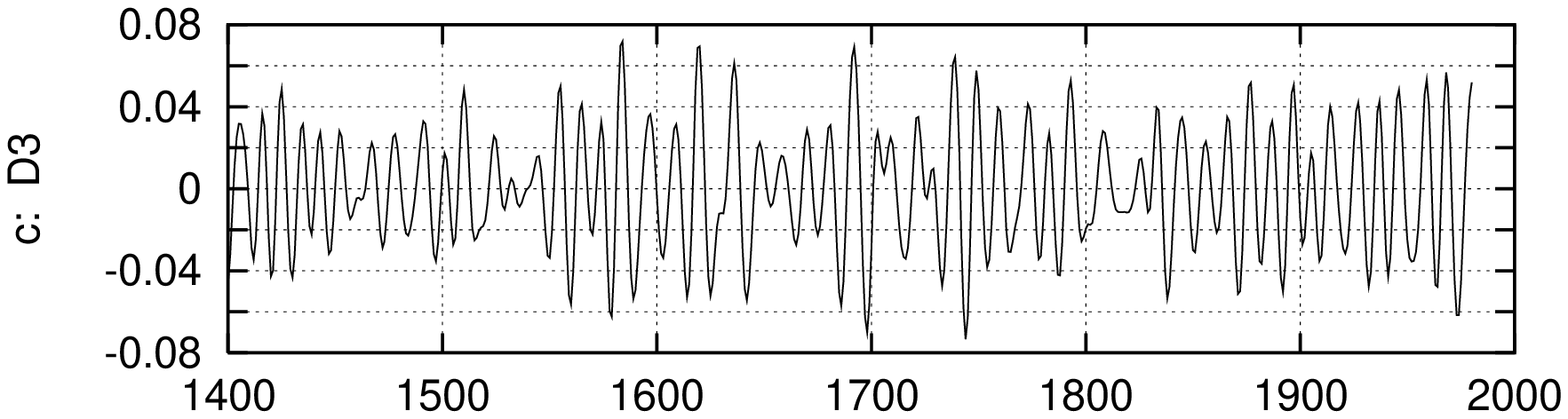,height=18cm,width=4cm,angle=-90}

\epsfig{file=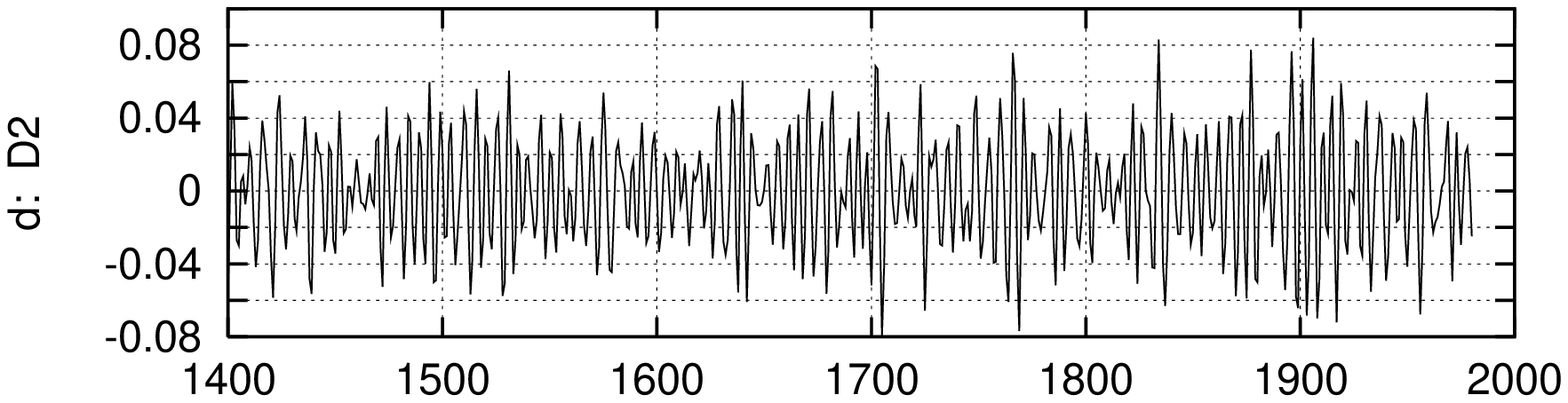,height=18cm,width=4cm,angle=-90}

\epsfig{file=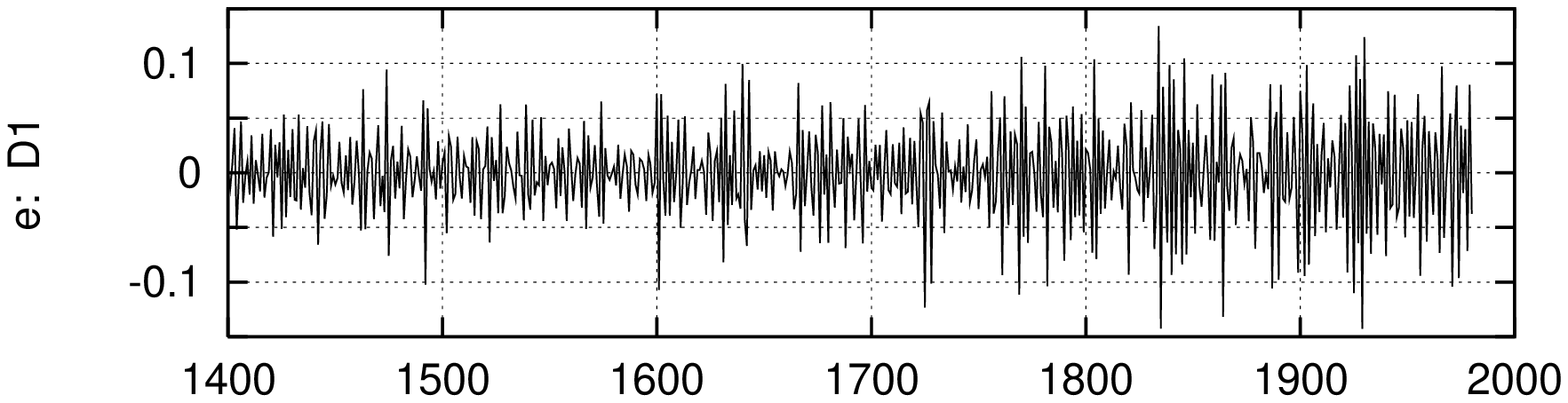,height=18cm,width=4cm,angle=-90}

\caption{Wavelet Multiresolution Analysis $J_0=4$ of the Mann temperature model; years 1400-1980. Fig. 4a shows the temperature model (dashed line) and the smooth average $S_4$ (solid line). Figs. 4b-e show the details $D_4$,  $D_3$, $D_2$, $D_1$.}

\end{figure}

\newpage

\begin{figure}

\epsfig{file=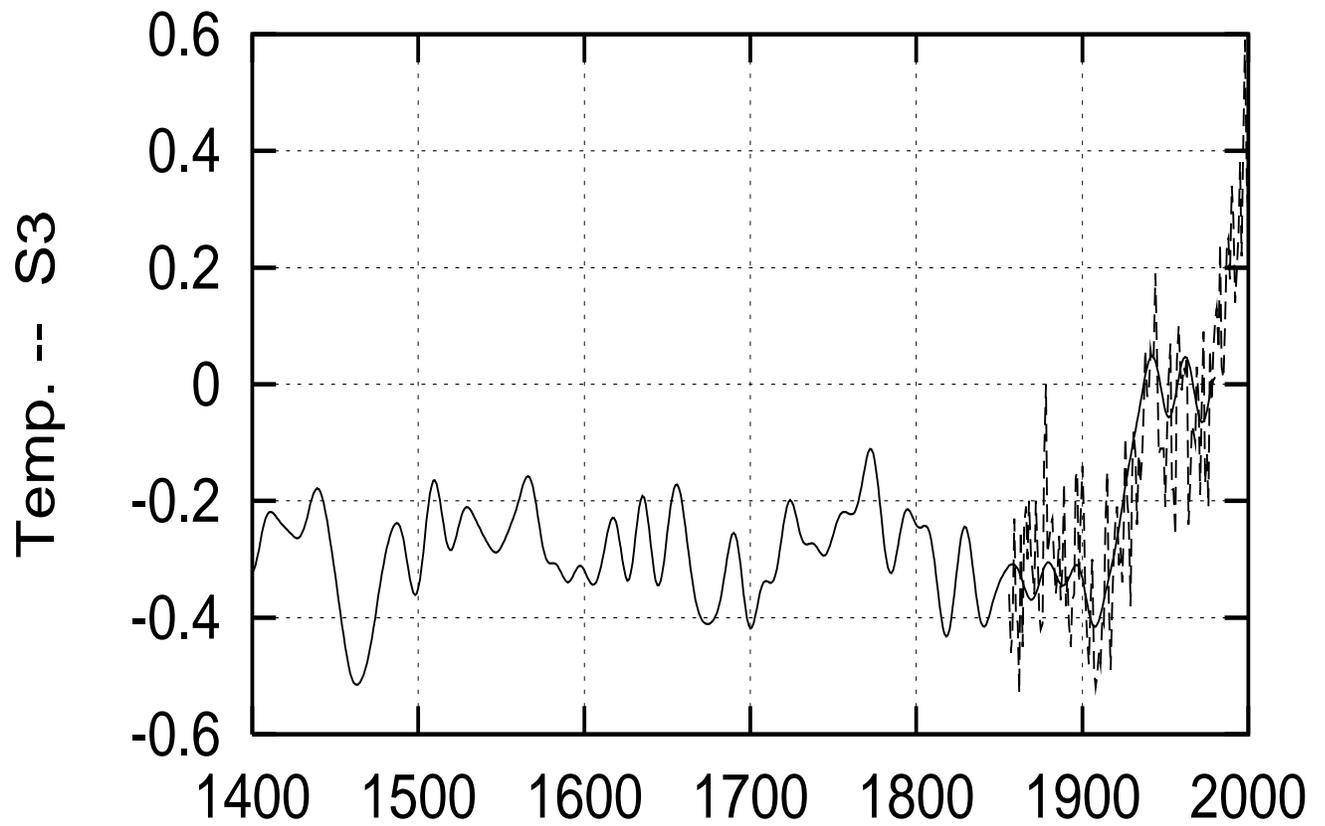,height=18cm,width=12cm,angle=-90}

\caption{ Mann temperature model: the smooth average $S_3$ that represent the most accurate smooth reconstruction of the temperature in the year 1400-1980 (solid line). The dashed line is the real temperature in the year 1856-2000.}

\end{figure}

\newpage

\begin{table}

  \begin{tabular}{|c|c|c|c|}

&r&r&r \\ 

          &  temp vs Briffa     &  temp vs Jones      &  temp vs Mann      \\ \hline

 data         &  0.23      & 0.56       &  0.86   \\ \hline

 S5         & 0.35          & 0.88      & 0.99      \\ \hline

 S4         &  0.34        &  0.84    &  0.99       \\ \hline

 S3         &  0.32         &  0.80    & 0.97      \\ \hline

 S2         &  0.27      &  0.72     & 0.93     \\ \hline

 S1         & 0.27          &  0.64    & 0.90     \\ \hline

 D5         & 0.82         & 0.76      & 0.96       \\ \hline

 D4         & 0.30           &  0.17    &0.66     \\ \hline

 D3         & 0.17          & -0.05       & 0.45    \\ \hline

 D2         & 0.30         & 0.10      & 0.62       \\ \hline

 D1         & 0.20         & 0.21      & 0.60    \\ \hline

  \end{tabular} 

\caption{Multiresolution Linear Correlation Analysis}

\end{table}

   \end{document}